\documentclass[12pt,a4paper]{article}

\newcommand{\arctanh}{\mathop{\mathrm{arctanh}}\nolimits}
\newcommand{\sech}{\mathop{\mathrm{sech}}\nolimits}
\newcommand{\csch}{\mathop{\mathrm{csch}}\nolimits}
\newcommand{\arcsech}{\mathop{\mathrm{arcsech}}\nolimits}
\newcommand{\arccsch}{\mathop{\mathrm{arccsch}}\nolimits}
\newcommand{\arcsec}{\mathop{\mathrm{arcsec}}\nolimits}

\title{Local Cosmology}

\author{Ll.\ Bel\\
\emph{Fisika Teorikoa, Euskal Herriko Unibertsitatea}, \\
\emph{P.K. 644, 48080 Bilbo, Spain} \\
\emph{e-mail: wtpbedil@lg.ehu.es}
}

\begin{document}
\maketitle

\begin{abstract} 

We define the concept of a Maximally symmetric
osculating space-time at any event of any given Robertson-Walker model.
We use this definition in two circumstances: i) to approximate any given
cosmological model by a simpler one sharing the same observational
parameters, i.e, the speed of light, the Hubble constant and the
deceleration parameter at the time of tangency, and ii) to shed some
light on the problem of considering an eventual influence of the overall
behaviour of the Universe on localized systems at smaller scales, or
viceversa. 

\end{abstract}

\section{Introduction}

Expressions such as Local cosmology, or Local cosmology of isolated
systems sound as a contradiction of terms because by cosmology we mean
the physics of that single isolated system that we call our Universe.
What our title reminds is that between the scale of the Universe as a
whole, which is certainly a more complex subject that any present or
future model will be able to contemplate, and that of much smaller
structures, there is room to be interested in those scales of
time and space which without spanning the whole universe include a
significant fraction of it.

In Sect.~2 we
propose to approximate a complex cosmological model by a simpler
one in the neighbourhood of any event at which the three fundamental
observational cosmological parameters are known. This approach mimics an
elementary geometrical construction. Consider a curve in a plane whose
Cartesian graph is a function $F(T)$. The parallel to the $T$ axis
through a point with coordinates $(T, F(T))$ is a very rough
approximation to the curve. If we want a better approximation we can draw the
tangent to the curve at the point. We still can do better by
drawing the osculating circle at the point. We will see that this
process suggests
to us to approximate a general Robertson-Walker model by either one of
five possible Robertson-Walker geometries with constant space-time curvature.

In Sect.~3 we compare the Robertson-Walker-like description of models 
with constant space-time curvature to their static one, emphasizing the problem 
of identifying the meaning of the radial coordinate in each case. We
conclude that the two descriptions do not correspond to different frames of reference 
but to different operational meanings of the radial coordinate.

In Sect.~4 we choose an appropriate expression for the energy-momentum
source of a Robertson-Walker geometry to be able to consistently derive
the evolution through time $T$ of space and space-time curvatures of the
corresponding osculating model.
 
In Sect.~5 we consider the problem of describing the perturbation of a
cosmological model by a single inhomogeneity, that we assume to be
spherically symmetric. Or equivalently the perturbation of a spherically
symmetric compact source that feels that spatial infinity is not flat
because it is modified by a Robertson-Walker behaviour. This problem has
been already considered in several papers (\cite{McVittie}-\cite{Gaite}). We take here a simplifying
view restricting ourselves to consider the problem in the context of
the dual description of Kottler's solution as a static or time dependent metric.

Our concluding remarks include some hints about how to deal with eventual
discrepancies between theory and observations that could be relevant to
the subject of this paper.
 

\section{Local osculating cosmological models}

Let us consider a Robertson-Walker Model that we shall write as:

\begin{equation}
\label{2.1}
dS^2=-dT^2+\frac{1}{c^2(T)}\left(\frac{dR^2}{1-
kR^2}+R^2d\Omega^2\right), \quad
d\Omega^2=d\theta^2+\sin^2\theta
d\varphi^2
\end{equation}
where $c(T)$ has dimensions of velocity and $k$ is the space curvature of the model.
We make thus explicit an obvious possible interpretation, \cite{Froth}, of
Robertson-Walker cosmological models as describing a continuous
sequence, labelled by a speed parameter constant\, \footnote{the word constant is here a language 
licence as one says the Hubble constant} $c(T)$, of static space-times where any
physical process whose evolution time-scale is fast compared to the
time-scale evolution of the effective speed of light $c(T)$ can be described. 

From a geometrical point of view the approximation that this
interpretation suggests amounts to approach a graph $c(T)$ by a parallel 
to the time axis taken at the time when the process is considered. In
this paper we want to improve on this idea of approximating a cosmological model
by a simpler one that in some precise sense is locally equivalent to it 
in some interval of $T$.

Let us consider the line-element (\ref{2.1}) at the time $T_{ref}$
when we start observing the local properties of the
space-time around our location. We shall assume that we have reset the clock
we use to: 

\begin{equation}
\label{2.2}
T_{ref}=0,
\end{equation} 
so that if we measure the effective speed of light at this time we obtain some
value
$c_0$ so that:

\begin{equation}
\label{2.3}
c(0)=c_0.
\end{equation}
What is usually called the scale factor is then the function:

\begin{equation}
\label{2.3.1}
F(T)=\frac{c_0}{c(T)}
\end{equation}
and from our point of view it is in fact a refractive index for which 
the speed of reference is $c_0$. 
We assume also that both the Hubble constant and the deceleration
parameter are known at $T=0$:

\begin{equation}
\label{2.4}
H_0=-\frac{\dot c_0}{c_0}, \quad q_0=-2+\frac{c_0\ddot c_0}{{\dot c_0}^2} 
\end{equation} 

We shall say that a Robertson-Walker model  with line-element:

\begin{equation}
\label{2.5}
dS^2=-dT^2+\frac{1}{\bar c^2(T)}\left(\frac{dR^2}{1-\bar k R^2}+
R^2d\Omega^2)\right)
\end{equation}
is maximally symmetric and osculating to the model defined by the
line-element (\ref{2.1}) at $T=0$ iff:

i) it has constant space-time curvature, i.e. the function $\bar
c(T)$ is a solution of the following two equations:

\begin{eqnarray}
\label{2.6}
{\dot{\bar c}}^2+\bar k \bar c^4&=&1/3\bar\Lambda \bar c^2 \\
{2\dot{\bar c}}^2-\bar c\ddot{\bar c}&=&1/3\bar\Lambda \bar c^2. 
\end{eqnarray}  
where $\bar k$ and $\bar\Lambda$ are respectively the space and the space-time 
curvature constants. 

ii) the two functions $c(T)$ and $\bar c(T)$ have in common their
value and the values of their first and second derivatives at $T=0$.

\begin{equation}
\label{2.7}
\bar c_0 =c_0, \quad  \dot{\bar c_0}=\dot c_0, \quad 
\ddot{\bar c_0}=\ddot c_0    
\end{equation}
ensuring thus that a line element and its osculating one have the same
observable values of $c_0$, $H_0$ and $q_0$.
  
From (\ref{2.3}), (\ref{2.4}) and the preceding conditions i) and ii) it follows
that:

\begin{equation}
\label{2.8}
\bar \Lambda=-3q_0H_0^2, \quad \bar k=-\frac{H_0^2}{c_0^2}(q_0+1)
\end{equation}
These formulas allow to distinguish five types of osculating
maximally symmetric space-times to a given Robertson-Walker model,
depending on the value of $q_0$:

1.- If $q_0=0$ the osculating space-time is Milne's model. In this
case one has:

\begin{equation}
\label{2.9}
\bar\Lambda =0, \quad \bar k=-\frac{H_0^2}{c_0^2}<0
\end{equation} 
and the function $\bar c(T)$ is:

\begin{equation}
\label{2.10}
\bar c(T):=\frac{1}{p(T+A)}, \quad A=\frac{1}{c_0p}
\end{equation}
where here and below the constant $A$ has been chosen such that
$c(0)=c_0$ and $p$ such that $p=\sqrt{\left|\bar k\right|}$.

2.- If $q_0=-1$ the osculating space-time is the proper de
Sitter model ($dS_0$). 
In this case one has:

\begin{equation}
\label{2.11}
\bar\Lambda =3H_0^2>0, \quad \bar k=0
\end{equation} 
and the function $\bar c(T)$ is:

\begin{equation}
\label{2.12}
\bar c(T)=c_0\exp(\lambda T), 
\end{equation}
where here and below $\lambda=\sqrt{\left|\bar\Lambda/3\right|}$. These two models
are not generic in the sense that unless one demands explicitly the
corresponding condition for the line-element (\ref{2.1}) one can not
expect to have either $k=0$ or $\Lambda=0$ exactly.

The generic models are the following:

3.- if $q_0<-1$ the osculating space-time is a positive
space-curvature de Sitter model ($dS_+$). In this case one has:

\begin{equation}
\label {2.13}
\bar k>0, \quad \bar \Lambda >0
\end{equation} 
and the function $\bar c(T)$ is:

\begin{equation}
\label{2.14}
\bar c(T)=\frac{\lambda}{p}\sech(\lambda(T+A)), \quad
A=\frac{1}{\lambda}\arcsech\left(\frac{pc_0}{\lambda}\right)
\end{equation}

4.- if $-1<q_0<0$ the osculating space-time is a negative
space-curvature de Sitter model ($dS_-$). In this case one has:

\begin{equation}
\label{2.15}
\bar k<0, \quad \bar\Lambda >0
\end{equation}
and the function $\bar c(T)$ is:

\begin{equation}
\label{2.16}
\bar c(T)=\frac{\lambda}{p}\csch(\lambda(T+A)), \quad
A=\frac{1}{\lambda}\arccsch\left(\frac{pc_0}{\lambda}\right)
\end{equation}

5.- if $q_0>0$ the osculating space-time is 
known as anti de Sitter model (AdS). In this case one has:

\begin{equation}
\label{2.17}
\bar k<0, \quad \bar\Lambda <0
\end{equation}
and the function $\bar c(T)$ is:

\begin{equation}
\label{2.18}
\bar c(T)=\frac{\lambda}{p}\sec(\lambda(T+A)), \quad
A=\frac{1}{\lambda}\arcsec\left(\frac{pc_0}{\lambda}\right)
\end{equation}


\section{Interpretation of the static description}
  
As it is well-known the line-elements corresponding to these five
models admit local time-like integrable Killing fields and therefore
they can be brought with an appropriate coordinate transformation to
the static form:

\begin{equation}
\label{3.1}
ds^2=-(1-\frac{\bar\Lambda}{3c_0^2}r^2)dt^2
+\frac{1}{c_0^2}(1-\frac{\bar\Lambda}{3c_0^2}r^2)^{-1}dr^2+
\frac{1}{c_0^2}r^2d\Omega^2
\end{equation}
We give below the corresponding coordinate transformations from this
static form to the line-elements corresponding to the five functions
$\bar c(T)$ listed in the preceding section. The $r$ coordinate
transformation is in every case:

\begin{equation}
\label{3.2}
r=\frac{Rc_0}{\bar c(T)}
\end{equation}  
while the corresponding time transformations are:

1.- For Milne's space-time:

\begin{equation}
\label{3.3}
t=\sqrt{1+p^2R^2}(T+A)
\end{equation}

2.-For the proper de Sitter's $dS_0$ space-time:

\begin{equation}
\label{3.4}
t=T-\frac{1}{2\lambda}\ln\left(1-\frac{\lambda^2R^2}{c(T)^2}\right)
\end{equation}

3.- For de Sitter's $dS_+$ model:

\begin{eqnarray}
\label{3.5}
t=\frac{1}{\lambda}\arctanh\left(\frac{(1+pR)\tanh(\frac12(\lambda(T+A))}
{\sqrt{1-p^2R^2}}\right)\\
+\frac{1}{\lambda}\arctanh\left(\frac{(1-pR)\tanh(\frac12(\lambda(T+A))}
{\sqrt{1-p^2R^2}}\right)
\end{eqnarray}

4.- For de Sitter's $dS_-$ model:

\begin{eqnarray}
\label{3.6}
t=\frac{1}{\lambda}\arctanh\left(\frac{pR+\tanh(\frac12(\lambda(T+A))}
{\sqrt{1+p^2R^2}}\right) \\
+\frac{1}{\lambda}\arctanh\left(\frac{-pR+\tanh(\frac12(\lambda(T+A))}
{\sqrt{1+p^2R^2}}\right)
\end{eqnarray}

5.-And finally for the anti de Sitter model:

\begin{eqnarray}
\label{3.7}
t=\frac{1}{\lambda}\arctan\left(\sqrt{1+p^2R^2}\tan(\frac12(\lambda(T+A))
+pR\right) \\
+\frac{1}{\lambda}\arctan\left(\sqrt{1+p^2R^2}\tan(\frac12(\lambda(T+A))
-pR\right)
\end{eqnarray}

The dual description of any model with constant space-time curvature
either in a static Killing frame of reference or in a
Robertson-Walker time-dependent conformal Killing frame of reference
raises a problem of physical interpretation that we want to
illustrate with the consideration of a particularly simple scenario.
Let us consider the static description  of the Anti de Sitter model.
An elementary calculation  shows that a free test particle can
describe circles of any radius $r$ around any point with angular
velocity:

\begin{equation}
\label{3.8}
\frac{d\varphi}{dt}=\lambda
\end{equation}
in the plane $\theta=\pi/2$. On the other hand these circular orbits will be described from the
Robertson-Walker point of view as spiralling orbits such that:

\begin{equation}
\label{3.9}
R=\frac{\lambda r}{p}\sec(\lambda(T+A))
\end{equation}  
The mathematics is transparent. With different variables we have
different laws of evolution, but we physicists have to understand better
the problem. We have to know the meaning of $r$, $t$, $R$ and $T$ and
the meaning of the coordinate transformations (\ref{3.2})-(\ref{3.7}) to reach
some understanding of the reality of the physical process we are
describing.

Let us assume that a test body, located at a point with radial
coordinate $r$, in the space-time and the frame of
reference described by (\ref{3.1}) has initial velocity zero at some
instant $t$. An
elementary calculation shows that at the same instant it has a radial
acceleration:
 
\begin{equation}
\label{3.10}
\frac{d^2r}{dt^2}=\frac13\bar\Lambda(1-\frac{1}{3c_0^2}\bar\Lambda r^2)r \qquad
(\frac{dr}{dt}=0)
\end{equation} 
This shows that even when $r$ is infinitesimal it can not be
considered to be a distance that can be measured with a stretched
thread because the origin $r=0$ being arbitrary such an interpretation 
is in contradiction with the homogeneity of the
space-time described by (\ref{3.1}). On the contrary if we consider 
the same
solution in the frame of reference described by (\ref{2.5}) we obtain:

\begin{equation}
\label{3.11}
\frac{d^2R}{dT^2}=0 \qquad
(\frac{dR}{dT}=0)
\end{equation} 
which is consistent with the interpretation of small values of $R$ as
distances measured with an stretched
thread. The meaning of $r$ that follows from (\ref{3.2}) is that small
infinitesimal values of this variable can be interpreted as Radar distances.

The interpretation of the variables $t$ and $T$ offers no problem
because they are both adapted synchronizations derived from proper
time along the same world-line $r=R=0$. 

\section{To follow through: an example}

Let us consider a particular Robertson-Walker model with line-element
(\ref{2.5}). We derive below convenient expressions of $\bar\Lambda(T)$ and 
$\bar k(T)$ 
corresponding to the osculating model at each time $T$ as explicit
functions of $c(T)$, $\rho(T)$ and $P(T)$. The latter being
respectively the mass density function of the model and the pressure
function. To do that we shall use the field equations in the
following form:

\begin{equation}
\label{6.1}
S_{\alpha\beta}+\Lambda g_{\alpha\beta}=8\pi GT_{\alpha\beta}
\end{equation}
where $S_{\alpha\beta}$ is the Einstein tensor of the line-element
(\ref{2.5}), $\Lambda$ is the proper global cosmological constant of
the model, with dimensions ${\hbox{Time}}^{-2}$ and $T_{\alpha\beta}$ is:

\begin{equation}
\label{6.2}
T_{\alpha\beta}=\rho(T)u_\alpha u_\beta 
+\frac{P(T)}{c(T)^2}(g_{\alpha\beta}+u_\alpha u_\beta)
\end{equation}
with $u_0=-1$ and $u_i=0$. Eqs. (\ref{6.1}) are then the usual equations with a small but important 
difference: the function $c(T)$ is a replacement for a constant $c_0$
considered to be a universal constant. Notice that this writing of
the field equations does not require any Einstein's gravitational
constant different from Newton's constant.

The explicit Einstein's equations (\ref{6.1}) reduce to the following
two:

\begin{equation}
\label{6.3}
3\frac{kc(T)^4+\dot c(T)^2}{c(T)^2}=8\pi G\rho(T)+\Lambda
\end{equation}

\begin{equation}
\label{6.4}
\frac{kc(T)^4+5\dot c(T)^2-2c(T)\ddot c(T)}{c(T)^2}=-8\pi
G\frac{P(T)}{c(T)^2}+\Lambda
\end{equation} 
 
Using the definitions (\ref{2.4}) for an arbitrary time:

\begin{equation}
\label{6.5}
H(T)=-\frac{\dot c(T)}{c(T)}, \quad q(T)=-2+\frac{c(T)\ddot
c(T)}{\dot c(T)^2} 
\end{equation}
we get solving the two algebraic equations with unknowns $H(T)^2$ and
$q(T)$:

\begin{equation}
\label{6.6}
H^2=-kc(T)^2+\frac13(8\pi G\rho(T)+\Lambda)
\end{equation}

\begin{equation}
\label{6.7}
q=\frac{4\pi G(3P(T)+\rho(T)c(T)^2)-\Lambda
c(T)^2}{c(T)^2(-3kc(T)^2+8\pi G\rho(T)+\Lambda}
\end{equation}
Using then the definitions (\ref{2.8}) with $c(T)$, $H(T)^2$ and
$q(T)$ instead of $c_0$, $H_0^2$ and $q_0$, we obtain finally:

\begin{equation}
\label{6.8}
\bar\Lambda(T)=\frac{-4\pi G(3P(T)+\rho(T)c(T)^2)+\Lambda c(T)^2}{c(T)^2}
\end{equation}

\begin{equation}
\label{6.9}
\bar k(T)=\frac{kc(T)^4-4\pi G(\rho(T)c(T)^2+P(T))}{c(T)^4}
\end{equation}
 
In \cite{Froth} we considered a model characterized by the appealing
equation of state:

\begin{equation}
\label{6.10}
P(T)=\frac13\rho(T)c(T)^2
\end{equation}
Using the field equations (\ref{6.1}), (\ref{6.2}) leads to a
particularly simple density dependence on $c(t)$:

\begin{equation}
\label{6.11}
\rho(T)=Bc(T)^4
\end{equation}
where B is a constant. The corresponding functions $\bar\Lambda(T)$ and
$\bar k(T)$ being:

\begin{equation}
\label{6.12}
\bar\Lambda(T)=\Lambda-8\pi G\rho(T), \quad 
\bar k(T)=k-\frac{16\pi G}{3c(T)^2}\rho(T)
\end{equation}

\section{Local cosmology of isolated systems}

Let us consider the Kottler solution \cite{Kottler}:

\begin{eqnarray}
\label{4.1}
ds^2&=&-(1-\frac{2GM}{c_0^2r}-\frac{\Lambda}{3c_0^2}r^2)dt^2+ \nonumber \\ 
&+&\frac{1}{c_0^2}(1-\frac{2GM}{c_0^2r}-\frac{\Lambda}{3c_0^2}r^2)^{-1}dr^2+
\frac{1}{c_0^2}r^2 d\Omega^2 
\end{eqnarray} 
which is a solution of Einstein's vacuum field equations including a
cosmological constant $\Lambda$. If $\Lambda=0$ it
becomes the proper Schwarzschild solution; and if $M=0$
then it reduces to the static descriptions of the space-times of constant
curvature we have considered in the preceding sections.

We consider the Kottler solution to be a convenient starting point to
present our ideas about what could be called the Local cosmology of
isolated systems. Roughly speaking this means assigning reality, up
to some approximation, to the asymptotically Robertson-Walker
description of the Kottler solution obtained by the coordinate
transformations (\ref{3.2})-(\ref{3.7}). 

Our approximation will consider only the linear approximation with respect
to $M$  and neglect every power of $\lambda r/c_0$, $\lambda R/c_0$ or
$\lambda/(c_0p)$ greater than 2. We shall assume also that $GM/c_0^2<<R$
and use it consistently with the preceding conditions neglecting,
for instance, terms as $(\lambda^2/c_0^2)(GM/c_0^2)R$.   

The final result of the calculation can be written as the line-element:

\begin{eqnarray}
\label{4.2}
dS^2&=&-\left(1-\frac{2GM\bar c(T)}{c_0^3R}\right)dT^2+ \nonumber \\
&+&\frac{1}{\bar c(T)^2}\left(\frac{1}{1-\bar kR^2}
+\frac{2GM\bar c(T)}{c_0^3R}\right)dR^2+
\frac{1}{\bar c(T)^2}R^2d\Omega^2 
\end{eqnarray}
where $\bar c(T)$ is each of the five functions listed in Sect 2. 
To reach this final result has required an additional  linear
time re-synchronization:

\begin{equation}
\label{4.3}
T \rightarrow T+GMQ(T)R/c_0^3 
\end{equation}
to make zero the coefficient $dTdR$. The corresponding function 
$Q(T)$ is for each case:

\begin{eqnarray}
\label{4.4}
&1.-& \quad Q(T)=4p \\
&2.-& \quad Q(T)=4\exp(\lambda T)\lambda/c_0 \\ 
&3.-& \quad Q(T)=2p\cosh(\lambda(T+A)) \\
&4.-& \quad Q(T)=2p\sinh(\lambda(T+A)) \\
&5.-& \quad Q(T)=-4p\sin(\lambda(T+A)) 
\end{eqnarray}

The Lagrangian derived from (\ref{4.2}) at the Newtonian
approximation is:

\begin{equation}
\label{4.5}
L=\frac12 F(T)^2 
({\dot R}^2+R^2({\dot\theta}^2+\sin^2\theta{\dot\varphi}^2))
+\frac{GM}{R}F(T)^{-1} 
\end{equation}
Notice that besides the variation of $G$ given by:

\begin{equation}
\label{4.6}
\tilde G=G F(T)^{-1} 
\end{equation}
the kinetic term gets also a time-dependent factor. And while the
Hamiltonian is obviously not conserved, because of the spherical symmetry,
the angular momentum:

\begin{equation}
\label{4.7}
P=F(T)^2R^2\sin^2\theta\dot\varphi
\end{equation}
is conserved, which means that the classical expression $R^2\sin^2\theta\dot\varphi$
it is not.
   
On the other hand the zero order approximation of the line-element
(\ref{4.2}) is the line-element (\ref{2.5}).

Our conclusion at this point is therefore that in the framework that
we have been proposing 

i) the equations of motion of a test body in the
field of a spherically symmetric source at the
lowest approximation are the equations derived from (\ref{4.5}):

\begin{equation}
\label{4.8}
\ddot R+2H\dot R=-\frac{GM}{R^2}F(T)^{-3}+
\frac{P^2}{R^3}F(T)^{-4}
\end{equation}

\begin{equation}
\label{4.9}
R^2\ddot\varphi+2R\dot\varphi(\dot R+HR)=0.
\end{equation}
assuming that the trajectory lies on the plane $\theta=\pi/2$.

ii) The Maxwell equations have to be written taking into account that
the space-time metric is (\ref{2.5}) and the transit time of light between 
two points of space has to be calculated taking into account that
light propagates along a medium with time-dependent index of refraction
$n=\bar F(T)$ as explained at the beginning of Sect. 2.

We believe that two points of view allow to implement this second
condition:

a.- The simplest one is to accept the minimally coupled vacuum 
Maxwell equations. In this case light will propagate along null
geodesics of (\ref{2.5}) and transit times should be calculated
accordingly. 

b.- The second possibility consists in sticking closer to the
interpretation of $\bar F(T)$, as suggested before for $F(T)$, i.e.:
\begin{equation}
\bar F(T)=\frac{c_0}{\bar c(T)}
\end{equation}
assuming accordingly
that the space-time trajectories of light are the null
geodesics of the space-time metric with coefficients \cite{Synge}:

\begin{equation}
\label{4.10}
\bar g_{\alpha\beta}=g_{\alpha\beta}+(1-\bar F^{-2})u_\alpha u_\beta, \quad
u_\alpha=-\delta_{0\alpha}
\end{equation}     
where $g_{\alpha\beta}$ are the metric coefficients of (\ref{2.5}). Or,
explicitly:

\begin{equation}
\label{4.10.1}
dS^2=-\frac{1}{\bar F(T)^2}dT^2+\frac{\bar F(T)^2}{c_0^2}\left(\frac{dR^2}{1-\bar k R^2}+
R^2(d\theta^2+\sin(\theta)^2d\varphi^2)\right)
\end{equation}

\section{Concluding remarks}

Let us assume for simplicity that using Newtonian theory, i.e.
the physics derived from (\ref{4.2}) with $\bar c(T)=c_0$, and
after taking into account every otherwise controlled perturbation, we
reach the conclusion that a discrepancy remains between  the theory and
observations\,\footnote{Two hypothetical discrepancies have been 
suggested recently \cite{Gutzwiller}-\cite{Anderson}}. It follows
from what we said in the previous section that we could try to explain the
discrepancy keeping in mind one or both of these two remarks:

i) Take into account that the variable $R$ is derived from telescopic 
observations while radar distances, say $R_{rad}$, measure round-trip travel-times of
light. To the second order approximation the relationship between these
two quantities is:

\begin{equation}
\label{4.11}
R_{rad.}=R+\frac{f}{2c_0}H_0R^2
\end{equation}
with $f=1$ if we choose option a.- to describe the propagation of light,
and $f=2$ if we choose option b.-.

ii) Consider the possibility of using (\ref{4.2}) and (\ref{2.5}) with
free parameters $\bar\Lambda$ and $\bar k$ and use consistently the
mechanics and the optics derived for them. Of course a single successful
explanation of a discrepancy using such freedom will not be reliable
until the same framework, with the same values of $\bar\Lambda$ and
$\bar k$, can be applied consistently to other aspects of the physics of the same
system.

The overall image that we feel legitimate to draw from this paper is
that together with the usual approach to Cosmology, that relies on models
that are supposed to describe the history of the Universe from its origin to its end,
it might be also worthwhile to describe it
piecewise considering different scales and different moments of
its history. Possible interesting space scales could start 
with the Earth-Moon and solar systems scales and beyond these reach those
of larger scales.

We want to suggest that each scale of this hierarchy of structures
could have its own Cosmology, so to speak, and have their own local
cosmological parameters $c_0$, $H_0$ and $q_0$. How much of it would
be just a reflection of the global behaviour of the Universe and how
much would be their own intrinsic properties remains an intriguing question.
We can even imagine that the global behaviour of the Universe is the
result of averaging effects and causes with roots at much smaller
scales. From this point of view the ideas and elementary results
presented in this paper could be considered as being more than mere
approximations. They could be seen as a more realistic way of trying
to understand the overall behaviour of our Universe.

\section*{Acknowledgments}

I gratefully acknowledge the position of visiting professor to the
UPV/EHU that I have been holding while this paper was being prepared. I
gratefully acknowledge the help provided by J.~M.~Aguirregabiria at an
early stage of the development of this paper as well as his later
control of its first draft. And I gratefully acknowledge stimulating
comments by A.~Feinstein and J.~M.~M.~Senovilla.


\begin{thebibliography}{99}

\bibitem{McVittie} G.C. McVittie, Mon. Not. R. Astron. Soc., {\bf 93},
325 (1933)

\bibitem{Dirac} P. A. M. Dirac, Proc. R. Soc. London, {\bf A365}, 19
(1979)

\bibitem{Gautreau} R. Gautreau, Phys. Rev. D, {\bf 29}, 198 (1984)

\bibitem{Gaite} A. Dom\'{\i}nguez and J. Gaite, Europhys. lett, {\bf 55}, 4, 458 (2001)
 
\bibitem{Froth} Ll. Bel, arXiv:gr-qc/9809051; arXiv:gr-qc/9905016  

\bibitem{Kottler} F. Kottler, Annalen Physik, {\bf 56}, 410 (1918)
\bibitem{Synge} J. L. Synge, Relativity: The General Theory, Ch. XI,{\S}
3, North Holland (1960)

\bibitem{Gutzwiller} M. C. Gutzwiller, Rev. Modern. Phys., {\bf 70}, 2,
Ch. XI, {\S} I, (1998)

\bibitem{Dumin} Yu. V. Dumin, arXiv:astro-ph/0302008  

\bibitem{Anderson} J. D. Anderson et al., Phys. Rev. D {\bf 65}, 082004 (2002)

\end{thebibliography}
\end{document}